\documentclass[3p,times,twocolumn]{elsarticle}
\usepackage{amssymb}
\usepackage{amsmath}%

\usepackage{hyperref}
\usepackage{lineno}
\usepackage{natbib}

\usepackage[utf8]{inputenc}
\usepackage[normalem]{ulem}
\usepackage{color}
\usepackage{xspace}

\newcommand{\atanh}{\ensuremath{{\rm atanh}}}
\newcommand{\jpsi}{\ensuremath{{\rm J}/\psi}}
\newcommand{\psiP}{\ensuremath{\psi(2 \rm S)}}
\newcommand{\Xpart}{\ensuremath{{\rm X}(3872)}}
\newcommand{\snn}{\ensuremath{ \sqrt{s_{\rm NN} } } }
\newcommand{\s}{\ensuremath{ \sqrt{s } } }

\newcommand{\PbPb}{\ensuremath{{\text{Pb--Pb}}}}
\newcommand{\KrKr}{\ensuremath{\text{Kr--Kr}}}
\newcommand{\pp}{\ensuremath{\rm pp}}
\newcommand{\TCF}{\ensuremath{T_{\rm CF}}}

\newcommand{\ccBar}{\ensuremath{\rm c\bar{c}}}

\newcommand{\der}{\ensuremath{ {\rm d}}}
\newcommand{\bT}{\ensuremath{\beta^{\rm s}_{\rm T}}}
\newcommand{\muB}{\ensuremath{\mu_{\rm B}} }

\newcommand{\pT}{\ensuremath{p_{\rm T}}}
\newcommand{\pt}{\ensuremath{p_{\rm T}}}
\newcommand{\mT}{\ensuremath{m_{\rm T}}}

\newcommand{\dsigdydpt}{\ensuremath{ \der ^{2}\sigma/\der \pt \der y}}

\newcommand{\BR}{\ensuremath{{\rm BR}(\Xpart \to \jpsi \pi^+ \pi^-)}}
\newcommand{\Kr}{\ensuremath{^{84}{\rm Kr}} }

\begin{document}

\newcommand{\corr}[1]{{\color[rgb]{1,0,0}{#1}}}
\renewcommand\sout{\bgroup \color{blue} \ULdepth=-.5ex \ULset}

\begin{frontmatter}



  \title{Transverse momentum distributions  of charmonium states with the statistical hadronization model}

  \author[1,2]{A.~Andronic}
  \author[2,3,4]{P.~Braun-Munzinger}
  \author[3]{M.~K.~K{\"o}hler\corref{cor1}}
  \cortext[cor1]{Corresponding author}
  \ead{markus.konrad.kohler@cern.ch}
  \author[2,5]{K.~Redlich}
  \author[2,3]{J.~Stachel}

  \address[1]{Westf\"{a}lische Wilhelms-Universit\"{a}t M\"{u}nster, Institut f\"{u}r Kernphysik, M\"{u}nster, Germany}
  \address[2]{Research Division and ExtreMe Matter Institute EMMI, GSI Helmholtzzentrum f\"{u}r Schwerionenforschung GmbH, Darmstadt, Germany}
  \address[3]{Physikalisches Institut, Ruprecht-Karls-Universit\"{a}t Heidelberg, Heidelberg, Germany}
  \address[4]{Institute of Particle Physics and Key Laboratory of Quark and Lepton Physics (MOE), Central China Normal University, Wuhan 430079, China}
  \address[5]{University of Wroc\l aw, Institute of Theoretical Physics, 50-204 Wroc\l aw, Poland}

  \begin{abstract}
    Calculations and predictions are presented within the framework of the statistical hadronization model for transverse momentum spectra of the charmonium states $\jpsi$, $\psiP$ and $\Xpart$ produced in nucleus-nucleus collisions at LHC energies. The results are confronted with available data and exhibit very good agreement by using particle flow profiles from state-of-the-art hydrodynamic calculations. For $\Xpart$ production in $\PbPb$ collisions we predict a transverse momentum distribution similar in shape to that for $\jpsi$  with a strong enhancement  at low transverse momenta and a production yield of about 1\% relative to that for $\jpsi$.  
  \end{abstract}

  
\end{frontmatter}

\section{Introduction}
\label{sec:Intro}

Ultra-relativistic heavy-ion collisions are widely used to investigate the evolution from the hot and dense phase of QCD with quarks and gluons as degrees of freedom, the quark-gluon plasma (QGP), towards
hadronization into color-neutral objects as degrees of freedom. Analysis of the  abundances of the resulting hadrons provides a reliable tool to characterize the phase boundary in the phase diagram of the strong interaction~\cite{BraunMunzinger:2008tz}.

The statistical hadronization model (SHM) is successfully used to predict and describe hadron abundances produced in relativistic nuclear collisions~\cite{Andronic:2017pug}. The underlying assumption is that at the latest at hadronization the fireball formed in such collisions is close to thermal equilibrium such that hadron yields can be characterized by a grand canonical partition function where baryon number, the $3$-component of the isospin, and strangeness are conserved on average and a rapid hadrochemical freeze-out takes place at the phase boundary. These assumptions connect the microscopic content of a heavy-ion collision with macroscopic attributes like the chemical freeze-out temperature $\TCF$ and the baryochemical potential $\muB$. 

In the light-flavor sector it is observed that, besides the frequently produced mesons and baryons, also more complex and loosely bound objects, such as light nuclei and even hypernuclei, along with their anti-particles, can be described within the SHM~\cite{Andronic:2010qu,Andronic:2017pug}. In Figure~\ref{fig:Fig1}, the primordial and total (anti-)particle yields computed within the SHM are shown as a function of the particle mass $m$ for a collision energy of $\snn = 2.76$~TeV\footnote{It should be noted, that throughout this letter, we use $c=k_{\rm B}= \hbar =1$.}. For $m \gg \TCF$, the primordial or thermal yields follow the relation $m^{3/2} \exp(-m/\TCF)$. The total yields include contributions from resonance decays and reproduce particle yields from data over nine orders of magnitude from the lightest mesons up to nuclei with mass number $A=4$.

A striking feature in the figure is the difference between the light-flavored loosely bound state hyper-triton $^{3}_{\Lambda}$H and the charmonium state $\jpsi$. Although their masses are very similar, the yield of $\jpsi$ is about three orders of magnitude larger. The origin of this enhancement is due to the statistical hadronization of charm quarks, formed in hard initial processes, where the number of charm quarks is conserved throughout the evolution of the collision. Technically, the conservation of charm quarks leads to a fugacity in the SHM for charmed hadrons~\cite{BraunMunzinger:2000px} which is, however, not a free parameter but determined by the measured  charm cross section. Charm quarks are not confined inside the QGP, thermalize within the QGP and hadronize at the QCD phase boundary into open and hidden charm hadrons. This SHM for charmed hadrons (SHMC) provides an excellent description of charmonium production~\cite{Andronic:2006ky,Andronic:2007bi,Andronic:2018vqh} without any new parameters and represents  compelling evidence, as demonstrated in Figure~\ref{fig:Fig1}, for this new production mechanism. A more detailed account of the SHMC is given below. 

Furthermore, a large degree of thermalization is observed in the spectra and the elliptic flow of $D$-mesons and their decay electrons~\cite{Abelev:2013lca,Adam:2016ssk}. A number of recent measurements have established the SHMC process  (sometimes dubbed `(re)generation') as the dominant production mechanism of $\jpsi$ in heavy-ion collisions at LHC energies~\cite{Abelev:2013ila,Adam:2016rdg,Acharya:2018jvc,Koehler:2018gmr}.  It is therefore appealing and important to extend the intriguing results of  $\jpsi$ production beyond yields to particle spectra and to more complex charmonium as well as open charm states to further investigate the SHMC mechanism. In the present publication we focus on charmonium states. Predictions for the open charm sector will be the subject of a future publication.

\begin{figure}[!t]
  \centering
  \includegraphics[width = 7.5 cm]{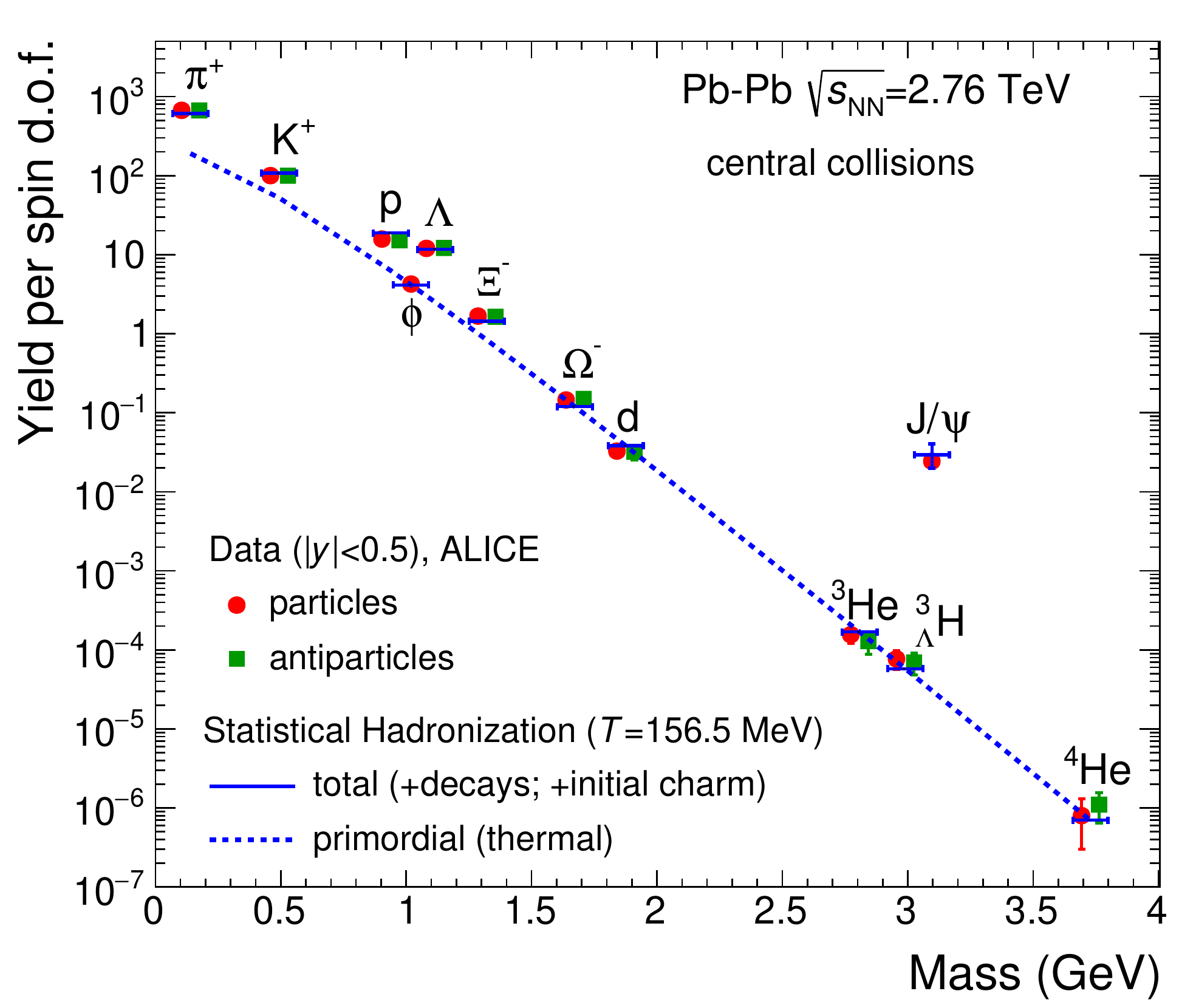}
  \caption{Primordial and total (anti-)particle yields, normalized to the spin degeneracy, as a function of mass calculated with the SHMC for $\PbPb$ collisions at $\snn = 2.76$~TeV and compared to data. See text for details.}\label{fig:Fig1}
\end{figure}
Loosely bound states such as $\psiP$ and, more dramatically,  the potential tetra-quark charmonium state $\Xpart$ are  of particular interest. Its observation by the Belle collaboration~\cite{Choi:2003ue} and the subsequent confirmation by the CDF~\cite{Acosta:2003zx}, D0~\cite{Abazov:2004kp}, and BaBar~\cite{Aubert:2004ns} collaborations showed that it is a narrow charmonium-like resonance and the close vicinity of the particle mass to the $D^0 \bar{D}^{*0}$ production threshold suggests that the particle could be a charm meson molecule with a very small binding energy~\cite{Tomaradze:2015cza}. At the LHC, the state $\Xpart$ was first observed by the LHCb collaboration~\cite{Aaij:2011sn}, which later also determined~\cite{Aaij:2015eva,Aaij:2013zoa} its quantum numbers, $J^{PC} = 1^{++}$. 

In addition to the $\Xpart$, the $\psiP$ is a natural choice when expanding the studies on the SHMC mechanism beyond the $\jpsi$. 

In this letter, we present calculations for the yields and transverse momentum spectra of the charmonium states $\jpsi$, $\psiP$, and $\Xpart$ for heavy-ion collisions at $\snn = 5$~TeV. Results will be presented for the current collision system $\PbPb$ and, in the case of $\Xpart$, also for $\KrKr$ collisions where much larger luminosities are possible at the LHC.

\section{Heavy quarks in the statistical hadronization model}
\label{sec:SHM}
In the SHMC it is assumed that charm quarks\footnote{In this letter we focus on charmonium, although the studies can be extended to any hadron species including charm and/or beauty content.} are produced in initial hard scatterings and that during the QGP phase the number of (anti-)charm quarks is conserved, i.e. the thermal production or annihilation are negligible at LHC energies~\cite{Andronic:2006ky}. The color charge of charm quarks is screened by the color-dense medium for $T \gtrsim \TCF = 156.5$~MeV and they do not form color-less bound states in the fireball volume $V$, where $\muB$ is consistent with zero for LHC energies, as determined by thermal fits, see~\cite{Andronic:2017pug}. The quarks thermalize in the QGP before the hadronization and rapid freeze-out at the phase boundary.

The (anti-)charm hadron densities computed in canonical statistical mechanics, $n_X^{{\rm th}}$, are anchored to the number of produced $\ccBar$ pairs, $N_{\ccBar}$, by a balance equation
\begin{equation}
  \begin{aligned}
    N_{\ccBar} =  \frac{1}{2} & g_c V \left( \sum_i  n^{{\rm th}}_{D_i} + n^{{\rm th}}_{\Lambda_i} + \cdots \right)  \\
    + & g_c^2 V  \left( \sum_i  n^{{\rm th}}_{\psi_i} + n^{{\rm th}}_{\chi_i} + \cdots \right) + \cdots,
  \end{aligned}
  \label{eq:balance}
\end{equation}
where the quantity $N_{\ccBar}$ is interpolated via FONLL~\cite{Cacciari:2012ny,Cacciari:2015fta} from charm cross-section measurements~\cite{Aaij:2013mga,Aaij:2015bpa,Aaij:2016jht,Acharya:2017jgo} in the corresponding rapidity region. Shadowing is taken into account when calculating $N_{\ccBar}$ for nucleus-nucleus collisions using rapidity-dependent measurements of the nuclear modification factor in proton-nucleus collisions of $D$-mesons~\cite{Aaij:2017gcy}, where interpolations, if necessary, are done via model calculations~\cite{Eskola:2009uj,Kovarik:2015cma,Kusina:2017gkz}. A canonical suppression factor, $I_1(g_c n_{{\rm oc}}^{{\rm th}}V)/I_0(g_c n_{{\rm oc}}^{{\rm th}}V)$, is applied to the open charm densities computed in grand canonical statistical mechanics to obtain $n_X^{\rm th}$ in eq.~\ref{eq:balance}. Here, $I_i$ with $i=\{0,1\}$ are modified Bessel functions and $n^{{\rm th}}_{{\rm oc}}$ are the thermal open charm meson densities. This correction factor gains importance for decreasing charm quark densities, i.e. $N_{\ccBar} \lesssim 1 $. In order to fulfill the balance equation (eq.~\ref{eq:balance}), the charm quark fugacity $g_c$ is needed. As explained before, it is not a free parameter but entirely defined by $N_{\ccBar}$ and $n_X^{\rm th}$. The parameter $V$ used in the equations above corresponds to the volume of one unit of rapidity and is determined in the statistical model fit of non-charm hadron yields~\cite{Andronic:2017pug}.

The density profile of the colliding nuclear species is taken into account by a core-corona approach. The `core' part represents the central region of the colliding nuclei where nucleons undergo many scatterings and are assumed to produce a QGP. The core fraction is normalized to the thermal yields. The `corona' area includes the nucleons in the outer region of the colliding nuclei. In the overlap zone with on average one or less collisions, no QGP formation is assumed. Rather the yields from the corona fraction are modeled by proton-proton ($\pp$) differential distributions scaled by the number of binary nucleon-nucleon collisions in the corona. The radius at which the nucleon density is not sufficient anymore to produce more than one inelastic scattering is taken from the nuclear charge density distributions. This density is found to be approximately at $10$~\% of the central nuclear density. To get a feeling for the sensitivity to this estimate, we also give the result for a value of  $20$~\%. The fraction of the core and the corona part is estimated by a Glauber simulation~\cite{Miller:2007ri}.

The calculation of the transverse momentum spectra is based on the following consideration: the charm quarks are assumed in local thermal equilibrium in the fireball formed in the collision. At hadronization, i.e. at $\TCF$, the charmonia states then `inherit' the random thermal motion of the charm quarks superimposed with the collective velocity of the expanding QGP fluid. This velocity is, hence, extracted from state-of-the-art viscous hydrodynamic modelling of light flavor observables. 

Specifically, this modelling is based on the (3+1)D viscous hydrodynamic simulation framework MUSIC~\cite{Schenke:2010nt} with IP-Glasma as initial condition\footnote{Here it is necessary to discriminate between the purpose of IP-Glasma and the previously mentioned shadowing. IP-Glasma affects the shape of the thermal contribution in the $\pT$-differential spectrum whereas shadowing affects the $\pT$-integrated $\ccBar$ cross section used to calculate the total number of produced $\ccBar$ pairs $N_{\ccBar}$ in a heavy-ion collision used in eq.~\ref{eq:balance}.}~\cite{Schenke:2012wb}. This framework does not assume boost-invariance. Rather the relevant parameters are adjusted to reproduce the charged particle multiplicity distributions measured by the ALICE collaboration as function of the pseudo-rapidity $\eta$~\cite{Abbas:2013bpa,Adam:2015kda}. The equation of state and the shear viscosity are parameterized from lattice QCD calculations~\cite{Dubla:2018czx}. The freeze-out hyper-surface is then taken at the chemical freeze-out temperature $T = \TCF$.

The results from these hydrodynamic simulations for the velocity profile $n$ and the transverse flow velocity $\bT$ are fed into a blast-wave function formulated for a Hubble-type expansion of the hyper-surface in a boost-invariant scenario as relevant for mid-rapidity data at LHC energy.  The result, inspired by earlier calculations reported in~\cite{Schnedermann:1993ws,Florkowski_book}, is:
\begin{equation}
  \begin{aligned}
     \frac{\der ^2 N}{\pT \der \pT \der y} &  \propto    \int_0^R r \der r \hspace{0.1 cm} \Bigg\{ \\
     \mT & \cosh\rho \hspace{0.1 cm}  K_1\left( \frac{\mT \cosh\rho}{T}\right) \hspace{0.1 cm} I_0\left(\frac{\pT \sinh\rho}{T} \right) \\
    -\pT & \sinh\rho \hspace{0.1 cm} K_0\left( \frac{\mT \cosh\rho}{T} \right)  \hspace{0.1 cm} I_1 \left( \frac{\pT \sinh \rho}{T} \right)  \Bigg\},
  \end{aligned}
  \label{eq:BW}
\end{equation}
where $\rho = \atanh(\bT (r/R)^{n})$ and $I_i$ and $K_i$ are modified Bessel functions with $i=\{0,1\}$. The transverse mass $\mT = \sqrt{m^2 + \pT^2}$  is obtained using the corresponding particle mass and the temperature is that of chemical freeze-out $T = \TCF$. 
This functional form (eq.~\ref{eq:BW}) is then used to model the thermal part of the spectrum and is normalized to the core fraction. To implement the boost-noninvariance discussed above in an approximate way for the description of data at forward rapidity, we adjust the flow velocity according to the (3+1)D hydrodynamic calculations, keeping otherwise the same functional form of the blast-wave description. 

We have tested numerically that using eq.~\ref{eq:BW} for large masses such as that of $\jpsi$ yields results very close to those obtained with the standard blast wave description of~\cite{Schnedermann:1993ws}.
The overall normalization factor is obtained from the SHMC as used in Figure~\ref{fig:Fig1}.

The shape of the corona fraction is modeled by a fit to the transverse momentum spectra measured in $\pp$ collision, given by
\begin{equation}
  f(\pT) = C \times \frac{\pT}{(1+(\pT/p_0)^2)^n},
  \label{eq:ppFit}
\end{equation}
where $C$, $p_0$ and $n$ are fit parameters.  

%
%
The recent measurement of ALICE is being used~\cite{Acharya:2019lkw} for the $\pT$-differential cross section of $\jpsi$ in $\pp$ collisions for mid-rapidity at $\snn = 5$~TeV.

\begin{figure}[t]
  \centering
  \includegraphics[width = 7.5 cm]{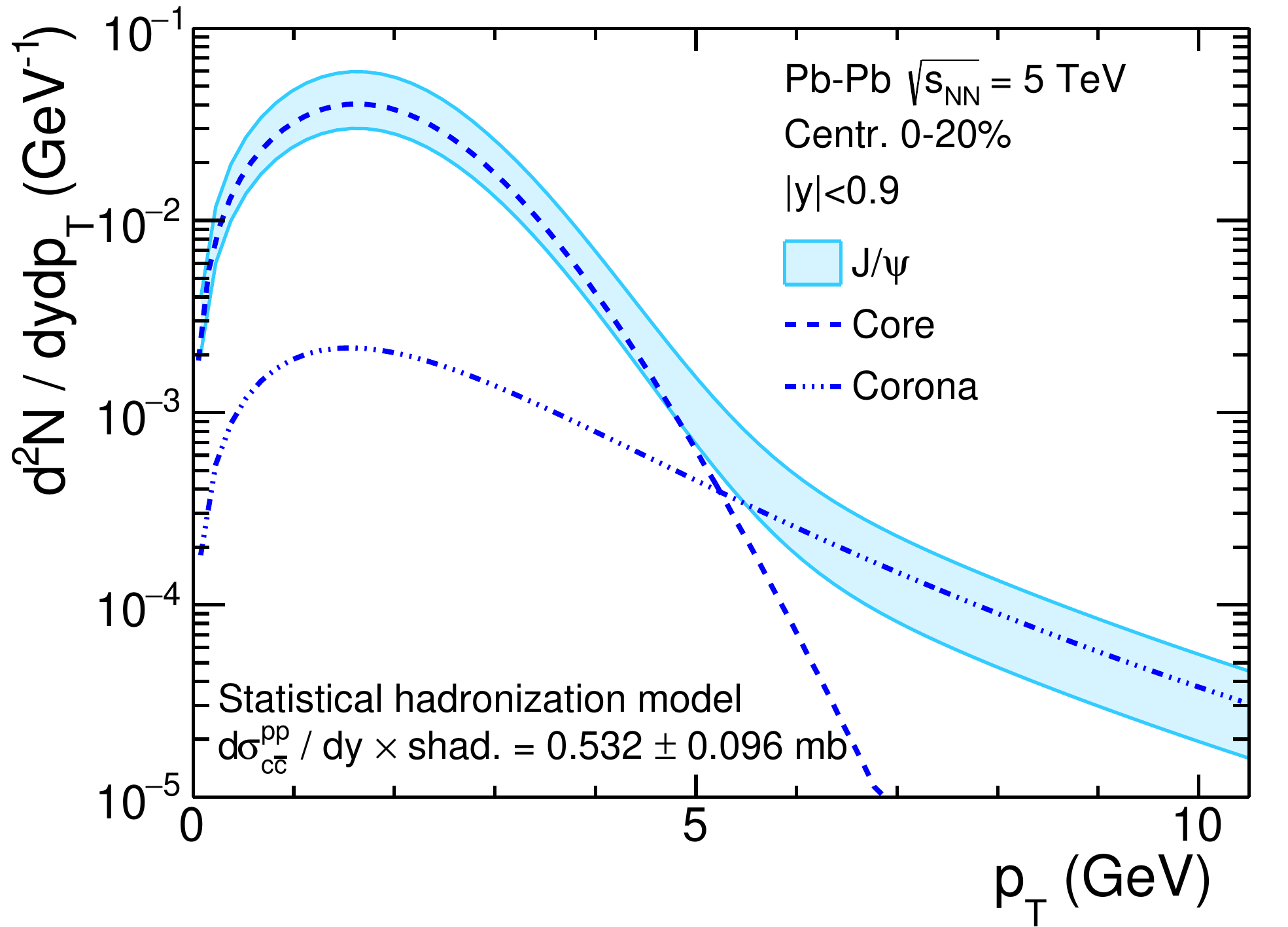}
  \caption{Transverse momentum spectrum at mid-rapidity $|y| < 0.9$ of $\jpsi$ for most central $\PbPb$ collisions at $\snn = 5$~TeV. The results are based on a charm cross section at mid-rapidity $|y|<0.9$ including shadowing as discussed above. In addition to the full spectrum calculation, the contributions for the thermal core part and the corona are shown. While at low $\pT$ the uncertainties are due to the charm cross section, at high $\pT$ the uncertainties come from the uncertainty of the corona thickness.}\label{fig:midy}
\end{figure}

%
%
In case of the $\psiP$, the $\pp$ cross section is taken from measurements at the corresponding collision energy and rapidity~\cite{Acharya:2017hjh}. The transverse momentum spectrum of $\psiP$ in $\pp$ collisions at low $\pT$ is not measured at mid- or forward rapidity. The shape is approximated by the $\jpsi$ transverse momentum spectrum from~\cite{Acharya:2017hjh}, which is consistent with the finding that the production ratio of $\psiP$ and $\jpsi$ are constant over a wide range in transverse momentum~\cite{Aad:2015duc}. Small differences can be expected for low transverse momentum~\cite{Aaij:2012ag,Acharya:2017hjh}, which should be negligible due to the dominant thermal contribution.

%
%
The $\Xpart$ $\pT$-differential cross section is measured at mid-rapidity for $10 < \pT / {\rm GeV } < 30$ in $\pp$ collisions at $\s = 7$~TeV~\cite{Chatrchyan:2013cld}. A universal collision energy scaling is assumed for the transverse momentum spectrum to extrapolate the spectrum from $\s = 7$~TeV to $ 5$~TeV for an estimate of the corona shape of the $\Xpart$ spectrum. `Hybrid' $\jpsi$ spectra in $\pp$ collisions are created from the ALICE and CMS results at $5$ and $7$~TeV to cover the low- and high-$\pT$ region simultaneously~\cite{Acharya:2019lkw,Sirunyan:2017mzd,Aamodt:2011gj,Chatrchyan:2011kc}. These spectra are fitted by the functional form eq.~\ref{eq:ppFit} and the ratio of the transverse momentum spectra parameterizations is used to weight the $\Xpart$ $\pT$ spectrum at $\s = 7$~TeV to estimate the shape of the spectrum at $\s = 5$~TeV. Finally, the $\Xpart$ cross section in $\pp$ collisions used for the determination of the corona fraction is estimated by an extrapolation from the $\pT$ range measured by CMS down to zero $\pT$ using eq.~\ref{eq:ppFit} and the corresponding $\pT$ spectrum at $\s = 5$~TeV.

\begin{figure}[t]
  \centering
  \includegraphics[width = 7.5 cm]{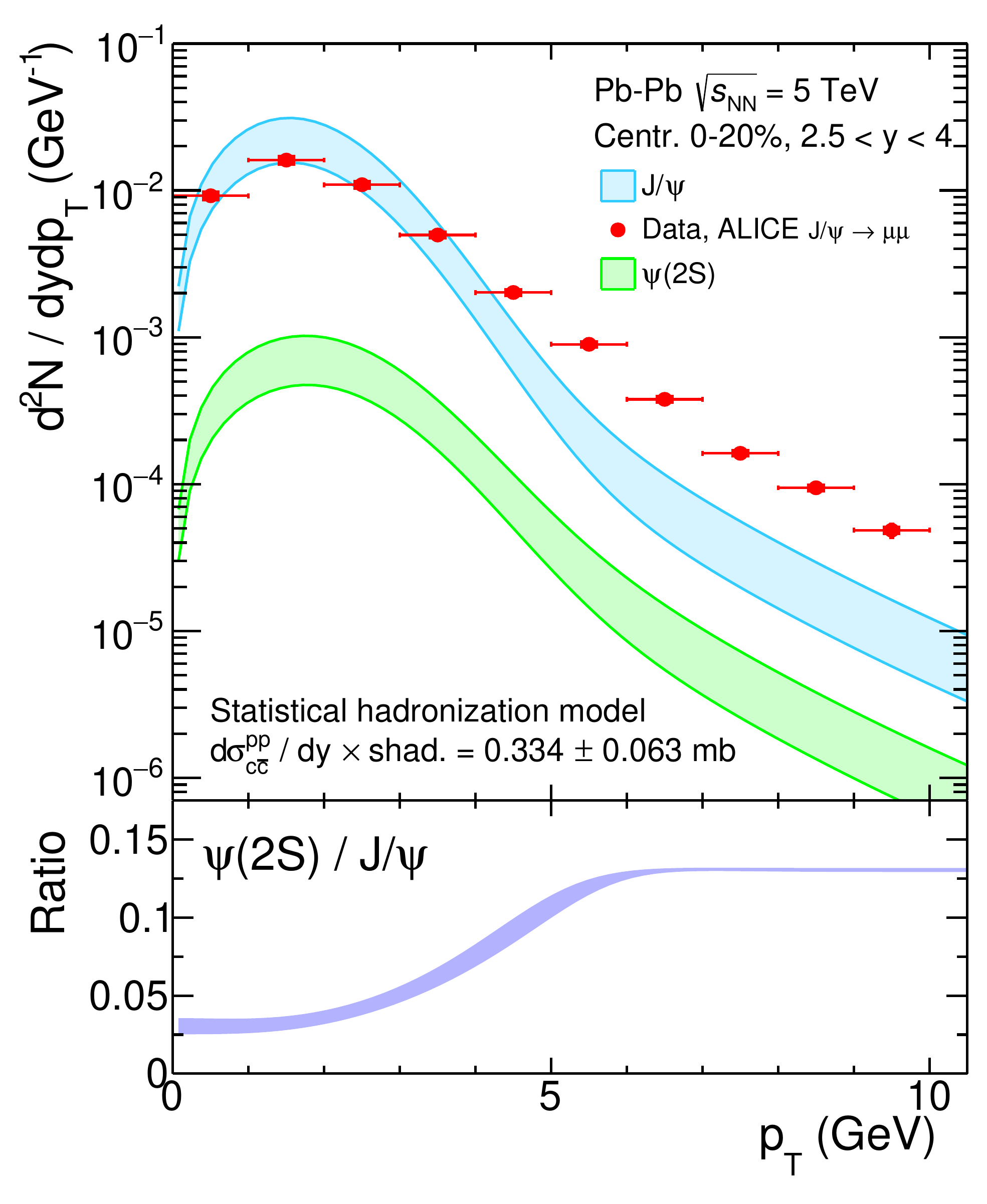}
  \caption{Upper panel: The transverse momentum spectrum for $\jpsi$ production in Pb-Pb collisions at $\snn = 5$~TeV obtained from ALICE data~\cite{Adam:2016rdg} in the forward rapidity region is compared to calculations from the SHMC. The value used for the charm cross section at forward rapidity is indicated in the figure. In addition, predictions for the $\psiP$ for the same charm cross section are shown for the centrality $0-20$~\%. Lower panel: Production ratio of $\psiP / \jpsi$.}\label{fig:forw}
\end{figure}

\section{Results}
In Figure~\ref{fig:midy}, the calculations are shown for the transverse momentum spectrum of the $\jpsi$ at mid-rapidity for $\PbPb$ collisions at $\snn = 5$~TeV. In addition to the full spectrum calculations the two components, thermal core fraction and corona, are plotted to visualize their contributions. One can see how dramatically different in shape the predicted $\pT$ spectrum in the hadronizing QGP is due to the large fugacity, see eq.~\ref{eq:balance} and  the strong collective transverse expansion. In minimum bias $\pp$ collisions such  effects are expected to be very small. 

In the upper panel of Figure~\ref{fig:forw}, calculations for the transverse momentum spectrum of $\jpsi$ in $\PbPb$ collisions at a collision energy of $\snn = 5$~TeV are compared to the measurement  by the ALICE collaboration at forward rapidity~\cite{Adam:2016rdg}. The model describes the data for low $\pT$ values but falls below the data for $\pT \gtrsim 4.5$~GeV. This suggests additional production mechanisms such as $\jpsi$ production from gluon fragmentation in jets which could contribute  towards increasing transverse momentum. In addition, predictions for the $\psiP$ are shown for the same collision system and rapidity window for central collisions. The lower panel shows the production ratio of $\psiP / \jpsi$. 

In Figure~\ref{fig:X3872}, predictions for the $\Xpart$ transverse momentum spectrum at $\snn = 5$~TeV are shown for $\PbPb$ and $\KrKr$ collisions. The branching ratio $\BR$ is only known approximately, but can be constrained to a range between $3.2$\% and about $20$\%~\cite{PDG:2018}. All results are therefore shown as $\BR \times \dsigdydpt$, where the branching ratio is chosen to $10$\%. Heavy-ion data taking in LHC Run4 or beyond the long shutdown 4 of the LHC, i.e. after the year $2030$, might enable these measurements. We also studied $\Xpart$ production in collisions between lighter beams such as  $\Kr$ nuclei to assess, in future experiments, the impact of potential increased luminosity {\it vs} reduced cross section. In this case, the yields in the $\KrKr$ collisions are found to be a factor $4-5$ lower than for $\PbPb$ collisions for $\pT \lesssim 5$~GeV.

\begin{figure}[t]
  \centering
  \includegraphics[width = 7.5 cm]{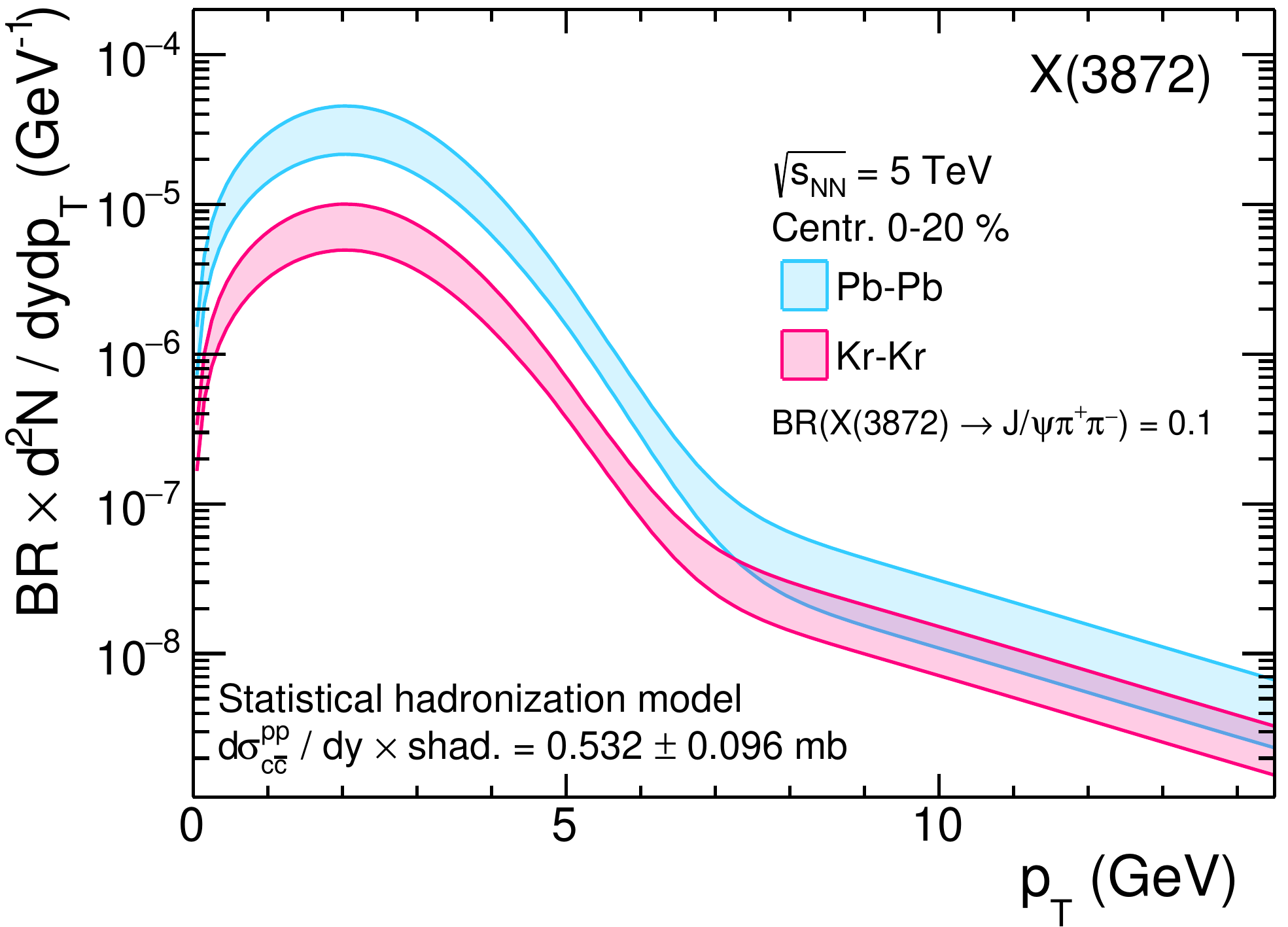}
  \caption{Predictions for the transverse momentum spectra of $\Xpart$ for $\PbPb$ and $\KrKr$ collisions at a collision energy of $\s = 5$~TeV for the centrality $0-20$~\%. The charm cross section corresponds to mid-rapidity at $\snn = 5$~TeV.}\label{fig:X3872}
\end{figure}

\section{Summary and conclusions}
Calculations and predictions using the  SHMC for the transverse momentum spectra of different charmonium states  are presented and, whenever measurements are available, confronted with data. The spectra calculations are based on the yields obtained from the SHMC as well as  a modified blast wave function with input from  state-of-the-art hydrodynamical calculations for the flow profiles. Without any further modifications or parameters a very good description of the measured $\jpsi$ transverse momentum spectra in the region $\pT \lesssim 5$~GeV is obtained. These results  further strengthen the evidence that charmonia are produced from deconfined charm quarks at the QCD phase boundary as their yields and spectra are established at the LQCD chiral crossover temperature.  

The comparison of the calculations with data for $\jpsi$ in $\PbPb$ collisions at $\snn = 5$~TeV at forward rapidity shows a good agreement for low $\pT$ but falls below the data for increasing momenta ($\pT \gtrsim 5$~GeV). This suggests additional production mechanisms beyond (re)generation at high charmonium momenta.

Predictions are presented for $\psiP$ production in $\PbPb$ collisions at $\snn = 5$~TeV at forward rapidity. Statistics collected in LHC heavy-ion data taking in LHC Run 3 might be sufficient for a comparison with data. These results will be important to understand open issues such as the possible presence, in the QGP, of color-less bound states. We further presented calculations for the transverse momentum spectrum of $\jpsi$ mesons produced in Pb-Pb collisions at mid-rapidity which can be tested soon.

Predictions for the production of $\Xpart$ for different centrality regimes of nuclear  collisions at $\snn = 5$~TeV are presented for $\PbPb$ and $\KrKr$ systems. The measurements might become available after the long shutdown 3 of the LHC and will provide a deeper understanding of the mechanism  of charmonium and exotica production at the QGP phase boundary and consequently about the QCD dynamics of deconfinement and the hadronization process.

\section*{Acknowledgment}
We thank Andrea~Dubla, Klaus~Reygers and Chun~Shen for fruitful discussions. 

This work is part of and supported by the DFG Collaborative Research Centre  ``SFB 1225 (ISOQUANT)". K.R. acknowledges support by the Polish National Science Centre under Maestro grant  DEC-2013/10/A/ST2/00106 and OPUS 2018/31/B/ST2/01663. K.R. also acknowledges partial support of the Polish Ministry of Science and Higher Education.






\bibliographystyle{elsarticle-num}
\biboptions{sort&compress}
\bibliography{biblio}

\end{document}